\documentstyle[12pt]{article}
\begin{document}
\setlength{\baselineskip}{5mm}
\newcommand{\ba}{\begin{array}}
\newcommand{\ea}{\end{array}}
\newcommand{\bea}{\begin{eqnarray}}
\newcommand{\eea}{\end{eqnarray}}
\newcommand{\eee}{\mbox{e}}
\newcommand{\triex}{\mbox{\hspace{3ex}}}
\newcommand{\ovl}{\overline}
\newcommand{\edoc}{\end{document}}
\newcommand{\pal}{\partial}
\newcommand{\sta}{\stackrel}
\newcommand{\lla}{\longleftarrow}
\newcommand{\lra}{\longrightarrow}
\newcommand{\ch}{\mbox{ch\,}}
\newcommand{\sh}{\mbox{sh\,}}
\newcommand{\th}{\mbox{th\,}}
\newcommand{\spsigma}{\ba[t]{c} \mbox{Sp} \vspace{-1ex} \cr
\mbox{$\scriptstyle{(\sigma)}$} \ea }
\newcommand{\spurab}{\ba[t]{c} \mbox{Sp} \vspace{-1ex} \cr
\mbox{$\scriptstyle{(a,b)}$} \ea }
%
\noindent{\large\bf
Free Fermions and Two-Dimensional Ising Model
}\vspace{4mm}

\noindent{V.\ N.\ Plechko $^a$
}\vspace{1mm}

\noindent{\small $^a$
Bogoliubov Laboratory of Theoretical Physics, Joint Institute
for Nuclear Research, 141980 Dubna, Moscow Region, Russia
}\vspace{4mm}

\setcounter{footnote}{1}
\footnotetext{\ Published in Proceedings of XXIII International Colloquium
on Group Theoretical Methods in Physics, July 31--August 5, 2000, JINR,
Dubna, Russia. Edited by  A.N.~Sissakian, G.S.~Pogosyan and L.G.~Mardoyan
(JINR Publ., Dubna, 2002), Vol.\ 2,\, p.~557--562.}

The two-dimensional (2D) Ising model (2DIM) is a one of only few nontrivial
solvable models in  statistical mechanics [1-18]. In its original
formulation, this is a discrete-spin lattice model for order-disorder
transition [1-7]. A remarkable feature, however, is that 2DIM admits as
well a field-theoretical interpretation in terms of free Majorana fermions
on a lattice [8-13,16,17]. This significantly simplifies the analytics of
the 2DIM and provides a new insight into the physical nature of the
problem. In this contribution, we shortly comment on some aspects of a
simple non-combinatorial fermionic approach to 2DIM based on the use of the
integrals over the anticommuting (Grassmann) variables and the
mirror-ordered factorization ideas for the 2DIM density matrix. A more
detailed discussion on related subjects is also given in [17]. Let us start
with a generalized formulation of the 2D Ising model, assuming arbitrary
inhomogeneous distribution of the bond coupling parameters over a
rectangular lattice net. The Ising spins, $\sigma_{mn}= \pm1$, are disposed
at the lattice sites, $mn$, with $m,n=1,...,L$ running in the horizontal
and vertical directions, respectively. At final stages, $N =L^2 \to\infty$.
The hamiltonian is:
\bea
-\beta\,H(\sigma)=\sum\limits_{m=1}^{L}\sum\limits_{n=1}^{L}\; \Big[\;
b_{m+1n}^{\,(1)}\,\sigma_{mn}\sigma_{m+1n}+ b_{mn+1}^{\,(2)}\, \sigma_{mn}
\sigma_{mn+1}\,\Big]\,,
\label{ham1}\;\; 
\eea
where $b_{mn}^{\,(\alpha)} =\beta J_{mn}^{\,(\alpha)}$ are dimensionless
bond coupling parameters, $\beta =1/kT$ is the inverse temperature. The
free boundary is assumed:  $\sigma_{L+1n}=0,\;\; \sigma_{mL+1}=0$. The
partition function is: $Z=\Sigma\,\exp\,(-\beta H)$, where the sum is taken
over the all possible spin configurations provided by $\sigma_{mn}=\pm1$ at
each site.  Noting an identity for a typical Boltzmann weight, $\exp\,(b\,
\sigma \sigma') =\cosh b +\sigma\sigma'\,\sinh b\,$, which readily follows
from $(\sigma\sigma')^{\,2} =1$, we come to the reduced partition function:
\bea
Q=\spsigma\Big\{\,\prod\limits_{m=1}^{L} \prod\limits_{n=1}^{L}
(\,1+t_{m+1n}^{\,(1)} \sigma_{mn} \sigma_{m+1n})\,(1+t_{mn+1}^{\,(2)}
\sigma_{mn}\sigma_{mn+1})\,\Big\}\,, \;
\label{qss1} \;\; 
\eea
where $t_{mn}^{\,(1,2)}=\tanh\,b_{mn}^{\,(1,2)}$, and $\,\mbox{Sp}_{
(\sigma)}$ stands for a properly normalized spin averaging, such that
$\mbox{Sp}\,(1) =1$, $\,\mbox{Sp}\,(\sigma_{mn}) =0$ at each site.  We
intend to convert $Q$ into a Gaussian fermionic integral. The starting
point is the fermionic factorization of the local Boltzmann weights from
(\ref{qss1}) [11,17]. For the whole lattice, we introduce a set
of the totally anticommuting Grassmann variables, $\,a_{mn},\,
a_{mn}^{\,*},\, b_{mn},\, b_{mn}^{\,*}\,$, and write:
\bea
&& 1+t_{m+1n}^{\,(1)}\sigma_{mn}\sigma_{m+1n}
=\,\int\limits_{}^{}da_{mn}^{\,*}da_{mn}^{}\, \eee^{\,\textstyle\mathstrut
a_{mn}^{}a_{mn}^{\,*}}\,(1+a_{mn}\sigma_{mn})\,
(1+t_{m+1n}^{\,(1)}\,a_{mn}^{\,*}\sigma_{m+1n})\,,\;\;\;
\nonumber \\[-1.5ex]
&& 1+t_{mn+1}^{\,(2)}\sigma_{mn}\sigma_{mn+1}
=\int\limits_{}^{}db_{mn}^{\,*}db_{mn}^{}\, \eee^{\,\textstyle\mathstrut
b_{mn}^{}b_{mn}^{\,*}}\,(1+b_{mn}\sigma_{mn})\,(1+t_{mn+1}^{\,(2)}\,
b_{mn}^{\,*}\sigma_{mn+1})\,.\;\;\;\;\;
\label{fac1}  
\eea
In a conventional notation, the bond Boltzmann weights are now presented
as $A_{mn}^{}A_{m+1n} ^{\,*}$ and $B_{mn}^{}B_{mn+1}^{\,*}$, where the
separable factors (to be called shortly Grassmann factors) are to be
identified from (\ref{fac1}).  The identities (\ref{fac1}) readily follows
from the elementary rules of fermionic integration.\footnote{ \ Let us
remember that Grassmann variables are the purely anticommuting fermionic
symbols. Given a set of Grassmann variables, $a_{1},a_{2},\ldots, a_{N}$,
we have: $a_{i}a_{j}+ a_{j}a_{i} =0\,,\,a_{j}^{\,2}=0$.  The rules of
integration for one variable are: $\int da_j \cdot a_j=1\,,\,\int  da_j
\cdot1=0$ [8]. The Gaussian fermionic integrals of any kind are
expressible in terms of the determinant (Pfaffian) of the correspondent
matrix [8].} At the next stage, we keep in mind to group
together, over the whole lattice, the four factors with the same spin (the
same index $mn$) and to sum over $\sigma_{mn}=\pm1$ in each group of
factors, independently, thus passing to a purely fermionic expression for
$Q$. The four Grassmann factors with the same spin are:
\bea
\ba{lllr}
& A_{mn}^{\,*} =1+t_{mn}^{\,(1)}a_{m-1n}^{\,*}\sigma_{mn}^{}\,,
& A_{mn}^{} =1+a_{mn}^{}\sigma_{mn}^{}\,,
\\[1ex]
& B_{mn}^{\,*} =1+t_{mn}^{\,(2)}b_{mn-1}^{\,*}\sigma_{mn}^{}\,,
& B_{mn}^{} =1+b_{mn}^{}\sigma_{mn}^{}\,.\;\;\;
\label{aaa1}\;  
\ea\eea
These factors are coming by factorization of the four different bonds
adjacent to a given $mn$ site. The problem is, however, that the separable
Grassmann factors like (\ref{aaa1}) are, in general, neither commuting
nor anticommuting with each other. So, a special ordering of factors in
their global products is needed in order the elimination of spin variables
be really possible [11,17].\footnote{ \ For a more detailed comment on this
subject see [11,17]. The construction of the mirror-ordered representation
(\ref{faq1}) for the density matrix is the clue point of the solution of
the 2D Ising model in a zero magnetic field within given approach
[11,12,16,17]. The ordering problem is an obstacle, however, to solve the
3D Ising model. A comment on the 2D Ising model in a nonzero magnetic field
is given below. } In two dimensions this problem is solvable by means of
the mirror-ordered arrangements of Grassmann factors in the total products
of weights (\ref{fac1}) forming the density matrix in (\ref{qss1}) [11,17].
Thus, we write:\vspace{-2ex}
\bea
Q\,(\sigma)= \spurab
\sta{n}{\sta{\lra}{\prod\limits_{n=1}^{L}}}\,
\Big[\,\prod\limits_{m=1}^{L}\sta{\sta{m}{-\!\lra}\;\;\;}
{A_{mn}^{\,*}B_{mn}^{\,*}A_{mn}^{}}\cdot\prod\limits_{m=1}^{L}
\sta{\sta{m}{\lla\!\!-}}{B_{mn}^{}}\,\Big]\,, \;\;
\label{faq1} \;\; 
\eea
where $\mbox{Sp}_{(a,b)}$ stands for the diagonal Gaussian averaging
arising in (\ref{fac1}). At any given fixed $n$,  we find here the two
ordered $m$-products of Grassmann factors, which are then multiplied over
$n$ (the modes of ordering are shown by arrows). The averaging over
$\sigma_{mn}=\pm1$ is to be performed at the junction of the two
$m$-ordered products in (\ref{faq1}), for given $n$. At each step, we
average over $\sigma_{mn}^{}=\pm1$ the product $A_{mn}^{\,*}B_{mn}^{\,*}
A_{mn}^{}B_{mn}^{}$ at the junction. This results an even fermionic
polynomial, equivalent to the Gaussian exponential factor just
corresponding to the local term (the $mn$-term) in the second line of
(\ref{qab1}). The equivalence of the polynomial to the exponential can be
checked either by a direct calculation, taking into account the nilpotent
properties of fermions, or making use of the identities like (\ref{lee1}),
see below. By a repeating use of the same procedure, taking at each stage
the commuting exponential factor away from the junction, we finally
eliminate all spin variables in (\ref{faq1}) [11,17]. The partition then
appears in the form:
\bea
&& Q\;=\; \int
\prod\limits_{m=1}^{L}\prod\limits_{n=1}^{L}
da_{mn}^{\,*}da_{mn}^{}db_{mn}^{\,*}db_{mn}^{}\;
\exp\,\Big\{\;\sum\limits_{m=1}^{L}\sum\limits_{n=1}^{L}\,\Big[\,
a_{mn}^{}a_{mn}^{\,*}+b_{mn}^{}b_{mn}^{\,*}\,+
\cr\cr
&& \,+\,t_{mn}^{\,(1)}t_{mn}^{\,(2)}\,a_{m-1n}^{\,*}b_{mn-1}^{\,*}
+\,(t_{mn}^{\,(1)}a_{m-1n}^{\,*}+t_{mn}^{\,(2)}b_{mn-1}^{\,*})\,
(a_{mn}^{}+ b_{mn}^{})\,+a_{mn}^{}b_{mn}^{}\,\Big]\Big\}\;,
\label{qab1}\;\;\;  
\eea
which is a Gaussian fermionic integral. Equivalently, the 2DIM is
reformulated as a theory of free fermions on a lattice. The above
expression for $Q$ is exact, assuming free boundary conditions for
fermions: $a_{0n}^{\,*} =0\,,\, b_{m0}^{\,*}=0$. For further
transformations of the integral (\ref{qab1}) (reduction to two variables
per site; continuum limit; effects of disorder) see [16,17].

Let us now consider the homogeneous rectangular lattice, $t_{mn}^{\,(1)}
=t_1,\, t_{mn}^{\, (2)} =t_2$. The integral (\ref{qab1}) can be explicitly
evaluated in this case by passing to the momentum space for fermions. In
the momentum space, the integral becomes:
\bea
Q\;=\; \int \prod\limits_{p=0}^{L-1}\prod\limits_{q=0}^{L-1}
da_{pq}^{\,*}da_{pq}^{}db_{pq}^{\,*}db_{pq}^{}\;
\exp\,\;\Big\{\;\sum\limits_{p=0}^{L-1}\sum\limits_{q=0}^{L-1}\,
\Big[\,a_{pq}^{}a_{pq}^{\,*} + b_{pq}^{}b_{pq}^{\,*}
+ a_{pq}^{}b_{L-pL-q}^{}\,+
\nonumber \\
+\, t_{1}t_{2}\,\eee^{\,i\,\frac{2\pi p}{L}-
\,i\,\frac{2\pi q}{L}}\,a_{pq}^{\,*}b_{L-pL-q}^{\,*} +
(t_1\,\eee^{\,i\,\frac{2\pi p}{L}}\,a_{pq}^{\,*} +t_2\,\eee^{\,i\,
\frac{2\pi q}{L}}b_{pq}^{\,*})\,(a_{pq}^{} + b_{pq}^{})\,
\Big]\,\Big\}\;,\;\;\;\;\;\;\;\;
\label{qap1} \; 
\eea
where $a_{pq}^{}, a_{pq}^{\,*}, b_{pq}^{}, b_{pq}^{\,*}$ are the new
variables of the integration introduced by the standard Fourier
substitution with periodic boundary conditions (change of boundary
conditions can be viewed as a boundary approximation inessential in the
limit of infinite lattice).  After a proper symmetrization of the fermionic
sum (action) from (\ref{qap1}) with respect to the $p,q \leftrightarrow
L-p,L-q$ conjugation, the integral decouples into a product of the
elementary low-dimensional integral factors, $Q_{pq}^{\,2}$. These integral
factors, $Q_{pq}^{\,2}$, can be readily evaluated by the standard rules of
fermionic integration [12,17]. Finally, we obtain an explicit
solution for the squared partition function in the form:
\vspace{-1mm}
\bea
Q^{\,2}\,=\,\prod\limits_{p=0}^{L-1}\prod\limits_{q=0}^{L-1}
\,\Big[\,(1+t_{1}^{\,2})(1+t_{2}^{\,2})-2t_1(1-t_{2}^{\,2})\,
\cos\frac{2\pi p}{L}- 2t_2(1-t_{1}^{\,2})\,\cos\frac{2\pi
q}{L}\,\Big]\,.\;\;
\label{qff7} \;\; 
\vspace{-1mm}
\eea
This is the exact solution for $Q^{\,2}$ of the 2D Ising model in the
limit $L^{\,2}\to\infty$. The correspondent free energy per site,
$-\beta f_{Q} =\frac{1}{L^2}\log Q\,|\,_{L^2 \to\infty}\,$, follows
in the form:
\vspace{-1mm}
\bea
-\beta f_{Q} =\frac{1}{2}\int\limits_{0}^{2\pi}\int\limits_{0}^{2\pi}
\frac{dp}{2\pi} \frac{dq}{2\pi}\ln\,\Big[(1+t_{1}^{2})(1+t_{2}^{2})
-2t_{1}(1-t_{2}^{2})\cos p - 2t_{2}(1-t_{1}^{2})\cos q\,\Big]\,,\;\;
\label{ftt1}\;\; 
\vspace{-1mm}
\eea
while the true free energy per site, for $Z=\Sigma\,\exp\,(-\beta H)$,
can be recalculated from $Z =(2\cosh b_{1}\cosh b_2)^{L^2}Q$, and
one finds:
*\vspace{-1mm}
\bea
-\beta f_{Z} =\ln 2  +\, \frac{1}{2}\,\int\limits_{0}^{2\pi}
\int\limits_{0}^{2\pi}\frac{dp}{2\pi}
\frac{dq}{2\pi}\ln\,\Big[\cosh 2b_1\cosh 2b_2
-\sinh 2b_1\cos p -\sinh 2b_2\cos q\;\Big]\,,\;\;\;
\label{fzz1} 
*\vspace{-1mm}
\eea
which is the famous Onsager's result, see Eq.~(108) in [1]. An interesting
comment by L.~Onsager on the history of his remarkable solution can be
found in [7]. The method we have applied, however, significantly differs
from the original approach [1].  As it follows from the exact solution, in
the ferromagnetic case, the critical point is given by the condition:
$1-t_1-t_2 -t_1\,t_2=0$, or $\,\sinh 2b_1\cdot \sinh 2b_2=1$. At this
point, the ($p=0,q=0$) mode in (\ref{qff7})-(\ref{fzz1})
vanishes.\footnote{ \ In the continuum-limit formulation, the parameter
$\underline{m} =1 -t_1 -t_2 -t_1t_2$ plays the role of the Majorana mass of
fermions [16,17]. It is interesting, that the Majorana-Dirac structures in
2D Ising model can be explicitly seen already at the lattice level, after
the reduction of the integral (\ref{qab1}) to two fermionic variables per
site [16,17]. The zero mass corresponds to the critical point.} For the
isotropic rectangular lattice, $t_1=t_2=t$, the critical point is
$t_c=\tanh b_c =\sqrt{2}-1$, or $\,\sinh 2b_c =1$, with the inverse
critical temperature $b_c=(J/kT_c) =\frac{1}{2}\, \ln(1+\sqrt{2})
=0.440686$. The singularity in the specific heat near $T_c$ appears to be
logarithmic, $C/k \simeq A_{c}\,| \log\tau| \to \infty$, with $\tau \propto
|T-T_c|\,, \tau\to0$. For the isotropic lattice, the specific-heat critical
amplitude is $\,A_{c}=\frac{8}{\pi}\, b_{c}^{\,2} =0.494539$. For more
comments on the properties of the 2DIM in a zero magnetic field see
[1,6,17,18].

Let us now turn back to the factorized density matrix (\ref{faq1}). The
related identities, considered below, may be of interest with respect to
the 2D Ising model in a nonzero magnetic field, for which the analytic
solution is yet unknown. Let $L_1$ and $L_2$ be arbitrary linear forms in
Grassmann variables, then:  $(1+L_1)\,(1+L_2) =\exp\,(L_1L_2)\, (1+L_1
+L_2)$, where the nilpotent properties of fermions where taken into
account. The two Grassmann factors here are combined into a one Grassmann
factor accompanied by a Gaussian exponential, $\exp\,(L_1L_2) =1+L_1L_2$.
The resulting identity can be iterated further on, and one finds: \\*[-3ex]
\bea
\prod\limits_{\alpha=1}^{N}\sta{\alpha}{\sta{\lra}{
\Big(1+L_{\alpha}\Big)}} \,=\Big(1
+\sum\limits_{\alpha=1}^{N}L_{\alpha}\,\Big)\,
\exp\,\Big(\sum\limits_{\alpha=1}^{N}
\sum\limits_{\beta=\alpha+1}^{N}L_{\alpha}L_{\beta} \Big)\,,  \;\;\;
\vspace{-2ex}
\label{lee1}  
*\vspace{-1mm}
\eea
where $L_\alpha$ are arbitrary linear forms in Grassmann variables, they
also can include the spin variables as parameters [17]. In
(\ref{lee1}) we assume $L_{N+1}=0$, similar conventions are assumed in
(\ref{lww2})-(\ref{ace6}) below. It is of interest also to consider the two
mirror-ordered products of factors like (\ref{lee1}). They can be combined
as follows: \\*[-3ex]
\bea
&& \prod\limits_{\alpha=1}^{N}\Big(\sta{\alpha}{\sta{-\!\longrightarrow}{
1+L_{\alpha}^{\,(1)}}}\Big)\cdot\prod\limits_{\alpha=1}^{N}
\Big(\sta{\alpha}{\sta{\longleftarrow\!-}{1+L_{\alpha}^{\,(2)}}}\Big)
=\Big(\,1+\sum\limits_{\alpha=1}^{N}\Big(L_{\alpha}^{\,(1)}
+L_{\alpha}^{\,(2)}\Big)\Big)\,
\nonumber \\
&& \times\,
\exp\Big\{\sum\limits_{\alpha=1}^{N}\sum\limits_{\beta=\alpha+1}^{N}
\Big(L_{\alpha}^{\,(1)} -L_{\alpha}^{\,(2)}\Big)\,\Big(L_{\beta}^{\,(1)}
+L_{\beta}^{\,(2)}\Big)\,+\sum\limits_{\alpha=1}^{N}
L_{\alpha}^{\,(1)}L_{\alpha}^{\,(2)}\,\Big\}\,.
\label{lww2} \;\;\; 
\eea
Making use of (\ref{lee1})-(\ref{lww2}), the factorized density matrix
(\ref{faq1}) can be elaborated into a mixed spin-fermion Gaussian integral
yet before the averaging over the spin variables. First, we apply the rule
(\ref{lee1}) to convert the local product of three factors
$A_{mn}^{\,*}B_{mn}^{\,*} A_{mn}^{}$ from (\ref{faq1}) into a one Grassmann
factor and accompanying Gaussian exponential. The exponensial factor in
fact appears to be a part of the local exponential factor from (\ref{qab1})
arising by the averaging over $\sigma_{mn}^{}$ at the junction in
(\ref{faq1}). The spin variables disappear from the exponential because of
$\sigma_{mn}^{2} =1$.  Then we combine the arising two $m$-ordered products
(at given $n$) according to (\ref{lww2}) with respect to index $m$, and
apply the rule (\ref{lee1}) once again with respect to $n$. At the second
step, the second line of (\ref{qab1}) will be effectively completed.
Finally, we come to the following representation for the spin-fermion
density matrix (\ref{faq1}):
\bea
&& Q\,(\sigma) =\int D\,\exp\,\{S_0\,(a,b)
+S_{\,int}\,(a,b,\sigma)\,\}\,,
\label{ano2} \;\;\; 
\eea
where $S_{\,0}^{}\,(a,b)$ is the spin-independent part of the action, the
same as in (\ref{qab1}), so that the integral $\int D \exp\,(S_0\,(a,b))$
is precisely the integral (\ref{qab1}), while $S_{\,int}\,(a,b,\sigma)$ is
the spin-fermion part of the action, which explicit form is:\\[-2ex]
\bea
\exp\,\{S_{\,int}\,(\sigma)\}
=\exp\Big\{\,\sum\limits_{m=1}^{L}
\sum\limits_{n=1}^{L}\Big[\sum\limits_{p=m+1}^{L}
L_{mn}^{\,(3)}L_{pn}^{\,(4)}\sigma_{mn}\sigma_{pn}
+\sum\limits_{p=1}^{L}\sum\limits_{q=n+1}^{L}
L_{mn}^{\,(4)}L_{pq}^{\,(4)}\sigma_{mn}\sigma_{pq}
\Big]\Big\}\,,\;
\label{abu2}
\eea
where $L_{mn}^{(3)} =(t_{mn}^{(1)}a_{m-1n}^{\,*}+t_{mn}^{(2)}
b_{mn-1}^{\,*}+a_{mn}^{}-b_{mn}^{})$ and $L_{mn}^{(4)}
=(t_{mn}^{(1)}a_{m-1n}^{\,*} +t_{mn}^{(2)}b_{mn-1}^{\,*} +a_{mn}^{}
+b_{mn}^{})$. The prefactor like $1+L_0$, where $L_0$ is a linear form
in fermions, is dropped in (\ref{abu2}), since, effectively, $L_0=0$ under
the integral. It can be guessed that the averaging over the spins in
(\ref{abu2}) gives unity, as it is to be expected, assuming that there are
no any other additional spin-dependent factors in the density matrix. The
additional factors will appear in the case of a nonzero magnetic
field. The expression (\ref{abu2}) can be elaborated in such a way, that
the spin variables can be eliminated as well in the presence of a field.
This results, however, a theory with a nontrivial four-fermion integration,
see (\ref{ace6}). Schematically, the spin-fermion part of action in
(\ref{abu2}) is of the form $S=\sigma a A a^{*}\!\sigma$, where $A$ is some
nonlocal matrix. Introducing auxiliary fermionic fields at each site,
$c,c^{\,*}$, the action can be transformed as follows:  $S=\sigma aAa^{\,*}
\sigma \to cc^{*} +\sigma\, (aAc +c^{*}a^{*}) \to cA^{-1\,T}c^{*}
+\sigma\,(ac+c^{*} a^{*})$. In the resulting action, the spin variables are
coupled to fermions linearly, so the averaging over the spins is not a
problem. Thus, introducing auxiliary fermionic fields,
$c_{mn},\,c_{mn}^{\,*}$, after some transformations, for the spin-dependent
part of action (\ref{ano2}) we find:\\[-3ex]
\bea
&&\exp\,\{S_{\,int}\,(\sigma)\}
=\int\prod\limits_{m=1}^{L}\prod\limits_{n=1}^{L} dc_{mn}^{\,*}dc_{mn}^{}
\exp\Big\{\sum\limits_{m=1}^{L}\sum\limits_{n=1}^{L}\,[\,
c_{mn}^{}c_{mn}^{\,*}+\sigma_{mn}^{}\,c_{mn}^{\,*}a_{mn}^{\,*}\,+\;\;\;
\nonumber \\
&&\,+\,\sigma_{mn}^{}a_{mn}^{}\,(c_{m+1n}^{}+ \ldots+ c_{Ln}^{})\,]
+\,\sum\limits_{n=1}^{L}\,c_{\underline{0}n}\,(c_{\underline{0}n+1}
+\ldots+c_{\underline{0}L})\,\Big\}\,,\;
c_{\underline{0}n} =\sum\limits_{m=1}^{L}\,c_{mn}\,.
\;\;\;\;\;\mbox{ \ } 
\label{ace1} \vspace*{-2mm}
\eea
Introducing new variables of integration by $c_{mn} \to$ $c_{mn}^{\,'}
{=}c_{mn}+c_{m+1n}+\ldots+ c_{Ln}^{}$, with the inverse transformation
$c_{mn}^{} =c_{mn}^{\,'} {-}c_{m+1n}^{\,'}$, we obtain another form of the
same identity:\\[-3ex]
\bea
&&  \exp\,\{S_{\,int}\,(\sigma)\}
=\int\prod\limits_{m=1}^{L}\prod\limits_{n=1}^{L} dc_{mn}^{\,*}dc_{mn}^{}
\exp\Big\{\sum\limits_{m=1}^{L}\sum\limits_{n=1}^{L}\,[\,c_{mn}^{}\,
\partial_{m}\,c_{mn}^{\,*}+\sigma_{mn}^{}\,
(a_{mn}^{}c_{m+1n}^{}\,+\, \mbox{ \ }
\cr
&& \,+\,c_{mn}^{\,*}a_{mn}^{\,*})\,]\,
+\sum\limits_{n=1}^{L}\,c_{1n}\,(c_{1n+1}+\ldots+c_{1L})\,\Big\}\,,
\;\;\;\partial_{m}c_{mn}^{\,*}=c_{mn}^{\,*}{-}c_{m-1n}^{\,*}\,.
\label{ace5} \;\;\;  
\eea
In the case of a nonzero magnetic field the additional Boltzmann factors
like $(1+t_{mn}^{\,(0)} \sigma_{mn}^{})$ are to be introduced into the
density matrix (\ref{faq1}), where $t_{mn}^{\,(0)} =\tanh h_{mn}$, and
$h_{mn} =\beta {\cal H}_{mn}$ is the dimensionless site-dependent magnetic
field. This corresponds to adding the terms like $(\ldots +h_{mn}
\sigma_{mn})$ into the hamiltonian (\ref{ham1}). The elimination of the
spin variables in the presence of a field in (\ref{ace5}) then
results:\\[-2ex]
\bea
\spsigma\!\exp\{S_{\,int}\,(\sigma)\}
=\!\int\!\prod\limits_{m=1}^{L}\prod\limits_{n=1}^{L}dc_{mn}^{\,*}
dc_{mn}^{}\exp\Big\{\sum\limits_{m=1}^{L}\sum\limits_{n=1}^{L}
\Big[\,c_{mn}^{}\partial_{m}c_{mn}^{\,*}
+t_{mn}^{\,(0)}(a_{mn}^{}c_{m+1n}+
\cr
+\,c_{mn}^{\,*}a_{mn}^{\,*}) + (1-t_{mn}^{\,(0)\,2}\,)\,
a_{mn}^{}a_{mn}^{\,*}c_{m+1n}^{}c_{mn}^{\,*}\,\Big]\,+
\sum\limits_{n=1}^{L}\,c_{1n}\,(c_{1n+1}+\ldots+c_{1L})
\,\Big\}.\, \;\;\;\; \label{ace6} \vspace*{-3ex}
\eea
Substituting (\ref{ace6}) into (\ref{ano2}), we obtain a purely fermionic
expression for the partition function of the 2DIM in a nonzero magnetic
field. The action that appears in (\ref{ace6}) is non-Gaussian, we deal
here with a system of interacting fermions. Therefore, the partition
function can not be straightforwardly calculated.\footnote{ \ For
conformal field theory analysis of 2DIM in a magnetic field at $T=T_c$ see
[14,15]. } There are few interesting features that can be observed in the
representations like (\ref{abu2})-(\ref{ace6}). We see that the long-range
fermionic correlations emerge in the 2D Ising model in a nonzero magnetic
field [17]. This statement can be formulated in few ways. In the identities
like (\ref{abu2}) or (\ref{ace1}), we find explicitly the fermions of one
sort to be coupled to the nonlocal sums of other fermionic variables
(accompanying spin variables may be assumed to be partly `frozen' by
switching on the field). In the representations like (\ref{ace5}) and
(\ref{ace6}), this is elaborated into the form in which extra lattice
fermions with zero mass (kinetic term $c_{mn}^{} \pal_{m}c_{mn}^{\,*}$) are
coupled to basic fermionic fields.  Under the elimination of the spin
variables in a nonzero field, on the other hand, a non-Gaussian quartic
term arises, which prevents the exact solution.  An approximation of the
Hartree-Fock type has been applied to analyze the critical properties of
2DIM in a field in [17]. The singular part of  the free energy near the
critical isotherm, $T=T_c,\, h\neq 0$, in the regime of `strong' magnetic
field, $\tau^{15/8}<\!\!< h<\!\!< 1$, has been conjectured in the
form:\\[-3ex]
\bea
-\beta\,f_{\rm\,sing}^{} =\frac{1}{2}\,\int
\frac{d^2p}{(2\pi)^2}\, \ln\Big(\bar{m}^2 + p^2 +
\frac{\lambda^2}{p^2}\Big)\,,\;\;\;
\label{fhh1}\;\; 
\vspace*{-1mm}
\eea
with $\bar{m}^2\propto\tau^2$, and $\lambda \propto h\,M(\tau,h)$, where
$M(\tau,h)$ is magnetization, $\tau\propto|T_c-T|\,,\,\tau\to 0$, and $h$
is the magnetic field. The case $\lambda=0$ exactly corresponds to the
singular part of the free energy in a zero field that follows from the
exact solution (\ref{ftt1})-(\ref{fzz1}). Within such approximation, the
singularity in the specific heat at the critical isotherm was found to be
logarithmic, $(C/k)_{\rm\, sing} =E_c\,|\,\ln h\,| \to\infty$, $h \to 0$,
with the amplitude $E_c=(8/15) \,A_c$, where $A_c$ is the thermal
specific-heat amplitude at the critical isobar ($T\neq T_c,\, h=0$) [17].
For isotropic lattice, $A_c=(8/\pi)\,b_{c}^{2} =0.\,494539$, and
$E_c=(64/15\pi)\,b_{c}^{2} =0.\,263754$, where $b_c =\frac{1}{2}
\ln(1+\sqrt{2})$. It might of interest to check these predictions by
Monte-Carlo experiments. In physical aspect, it seems also to be highly
desirable to clarify the mechanism of the spontaneous ordering in the 2D
Ising model in terms of fermions.

\newcommand{\etal}{{\em et al.}}
\setlength{\parindent}{0mm}
\vspace{5mm}
{\bf References}
\begin{list}{}{\setlength{\topsep}{0mm}\setlength{\itemsep}{0mm}%
\setlength{\parsep}{0mm}}

\item[1.]  L.~Onsager, Phys. Rev. {\bf 65}, 117 (1944).
\item[2.]  E.~W.~Montroll, R.~B.~Potts, and J.~C.~Ward, J. Math. Phys.
{\bf 4}, 308 (1963).
\item[3.] T.~D.~Schultz, D.~C.~Mattis, and E.~H.~Lieb, Rev. Mod. Phys.
{\bf 36}, 856 (1964).
\item[4.]
L.~P.~Kadanoff and H.~Ceva, Phys. Rev. B {\bf 3}, 3918 (1971).
\item[5.]
T.~T.~Wu, B.~M.~McCoy, C.~A.~Tracy, and E.~Barouch, Phys.~Rev.
B{\bf 13}, 316 (1976).
\item[6.]
K.~Huang, Statistical Mechanics (Wiley, New York, 1987).
\item[7.]
L.~Onsager,  In: Critical Phenomena in Alloys, Magnets
and Superconductors, R.~E.~Mills, E.~Ascher, and R.~I.~Jaffe, editors
(McGraw-Hill, New York, 1971), p. XIX-XXIV; p. 3--12.
\item[8.]
F.~A.~Berezin, Russ. Math. Surveys, {\bf 24}, No.~3, 1 (1969).
\item[9.]
S.~Samuel, J. Math. Phys. {\bf 21}, 2806 (1980).
\item[10.]
C.~Itzykson, Nucl. Phys. B {\bf 210} [FS6], 448 (1982).
\item[11.]
V.~N.~Plechko, Sov. Phys. Doklady, {\bf 30},\, 271 (1985).
\item[12.]
V.~N.~Plechko, Physica A, {\bf 152},\, 51\, (1988).
\item[13.]
C.~Itzykson and J.-M.~Drouffe,  Statistical Field Theory (Cambridge
University Press, Cambridge, 1989).
\item[14.]
A.~B.~Zamolodchikov, Adv. Stud. Pure Math. {\bf 19}, 641 (1989).
\item[15.]
G.~Delfino and G.~Mussardo, Nucl. Phys. B {\bf 455} (1995) 724.
\item[16.]
V.~N.~Plechko, Phys. Lett. A {\bf 239}, 289 (1998).
\item[17.]
V.~N.~Plechko, J. Phys. Stud. (Ukr)\, {\bf 3}, 312 (1999). A Special
Issue Dedicated to 90\,th Anniversary of Professor  N.~N.~Bogoliubov.
\item[18.]
B.~M.~McCoy, The 1999 Heineman Prize Address. Integrable Models in
Statistical Mechanics: The Hidden Field With Unsolved Problems. --
math-ph/9904003.
\end{list}
\end{document}